\documentclass[11pt,onecolumn]{IEEEtran}

\usepackage[utf8]{inputenc}
\usepackage[english]{babel}
\usepackage{authblk}
\usepackage{srcltx}
\usepackage{amssymb}
\usepackage{graphics}
\usepackage{psfrag}
\usepackage{epsfig}
\usepackage{subfigure}
\usepackage{array}
\usepackage{algorithm}
\usepackage{algorithmic}
\usepackage{makeidx} 
\usepackage{bbm}
\usepackage{hhline}
\usepackage{eufrak}
\usepackage{yfonts}
\usepackage{pifont}
\usepackage{footnote}
\usepackage{setspace}
\usepackage{amsfonts}
\usepackage{amsmath}

\usepackage{color}
\usepackage[square, comma, sort&compress, numbers]{natbib}
\usepackage{dsfont}

\begin{document}
\title{\huge{Union of Low-rank Subspaces Detector}}
\author[1]{Mohsen Joneidi}
\author[2]{Parvin Ahmadi}
\author[2]{Mostafa Sadeghi}
\author[1]{Nazanin Rahnavard}
\affil[1]{Department of Electrical Engineering and Computer Science, University of Central Florida}
\affil[2]{Department of Electrical Engineering, Sharif University of Technology}
\maketitle
\begin{abstract}
 The problem of signal detection using a flexible and general model is considered. Due to applicability and flexibility of sparse signal representation and approximation,  it has attracted a lot of attention in many signal processing areas. In this paper, we propose a new detection method based on sparse decomposition in a union of subspaces (UoS) model. Our proposed detector uses a dictionary that can be interpreted as a bank of matched subspaces. This improves the performance of signal detection, as it is a generalization for detectors. Low-rank assumption for the desired signals implies that the representations of these signals in terms of some proper bases would be sparse. Our proposed detector exploits sparsity in its decision rule. We demonstrate the high efficiency of our method in the cases of voice activity detection in speech processing.
\end{abstract}
\section{Introduction}
Sparse approximation techniques have found wide use due to their benefits and high flexibility in many applications in image and signal processing \cite{Elad1}\cite{Bruckstein2}. Sparse representation can efficiently extract most important features of a signal, so it provides very promising results in data compression \cite{RahmouneVF3}, de-noising \cite{Elad4}, blind source separation \cite{BSS5}, signal classification \cite{Mairal6}, and so on. The methods based on exploiting the signal sparsity have two main steps. First, an over-complete dictionary \cite{Elad1} is selected/learned according to the structural characteristics of the set of signals, and then the target signal is decomposed over the dictionary to obtain a compact representation. Representation in terms of a few designed/learned bases can accurately capture the signal structure characteristics, which in turn, leads to an improvement in the distinction between noise/interference and structured signals.

In some signal processing applications, the task is to detect the presence of a signal from its noisy measurements. For example, in speech processing, Voice Activity Detection (VAD) is performed to distinguish speech segments from non-speech segments in an audio stream. VAD plays a critical role on increasing the capacity of transmission and speech storage by reducing the average bit-rate \cite{VAD620527}.

Signal detection is an old problem in signal processing and there are some traditional signal detectors including energy detector, matched filter and matched subspace detector \cite{Scharf12}. Matched signal detector is the most basic framework for signal detection which needs a bank of matched signals to design a detector system. However, in many applications it is preferred to replace rank-1 signals by a multirank matched subspace \cite{Scharf12}. Matched subspace detector assumes the span of a subspace as the desired signals and rejects that part of signal which lies on the null-space of the assumed subspace. Generalized likelihood ratio test (GLRT) for matched subspace detector is the uniformly most  powerful invariant (UMP-invariant) statistic for detection \cite{Scharf12}. The subspace model for detection needs some bases as the span of desired signals which can be a set of fixed bases like discrete fourier transform (DFT) or data-dependent bases like principal component analysis (PCA). Although subspace model is more adaptive for signal analysis, it needs several parameters that must be either known or estimated. For example the set of bases spanning the desired signals, the coefficients of the bases, noise covariance and signal to noise ratio (SNR). Depending on the knowledge about different parameters, the optimum statistic is suggested in \cite{Scharf890324} for 4 situations. In the case of unknown coefficients, orthogonal projection of the observation is used to determine the coefficients of the contribution of each basis. The present paper assumes a more general model for signals in which considers a union of subspaces.

Sparsity has been exploited widely in detection purposes, e.g., abnormal event detection \cite{Ahmadi}, voice activity detection \cite{Teng24}, face detection \cite{Le7651}. A multi criteria detection based on intelligent switching between traditional detection and sparse detection is proposed in \cite{Byonghyo6146}. In these works, sparsity has been used to extract features or define a heuristic criterion for detection. Compressive detector is another application of sparsity for signal detection. It is able to detect signals only by using some measurements from the original samples while the performance is not degraded dramatically \cite{Duarte651} \cite{Davenport16}. The goal of compressive detector is to preserve the performance of detector the same as the original detector. In this paper we use sparsity from a different point of view. The traditional detectors are generalized to consider sparsity on the optimum decision rule and a new trade-off is suggested between sparsity (rank of a subspace) and error of projection (distance to a low-rank subspace).

In this paper, we propose a new signal detection method based on the union of low-rank subspaces (ULRS) model \cite{Lu7} \cite{Liu8}. This model is able to reveal intrinsic structure of a set of signals. The proposed detector is a generalized version of traditional detectors. In other words, imposing a union of rank-1 subspaces model for desired signals yields nothing other than the traditional matched filter banks. We investigate our detector from different points of views in order to show relation between our method and other classical detectors. We also derive a robust version of the proposed detector in order to provide robustness against outliers and gross errors. We provide theoretical investigations as well as experimental results on VAD.

The rest of the paper is organized as follows. Section 2 provides a brief background on sparse representation theory and basic concepts of detection theory. In Section 3 we describe our new signal detection method, study its performance and provide its robust version. Section 4 experimentally demonstrates the effectiveness of our proposed signal detection method. Finally, Section 5 concludes the paper with a summary of the proposed work.
\section{Theoretical Background and Review}
\subsection{Basic Theory of Sparse Decomposition}
Sparse decomposition of signals based on some basis functions has attracted a lot of attention during the last decade \cite{Elad1}. In this approach, one wants to approximate a given signal as a linear combination of as few basis functions as possible. Each basis function is called an atom and their collection is called a dictionary \cite{Mallat9}. The dictionary is usually over-complete, i.e., the number of atoms is much more than the dimension of atoms. Specifically, let $\boldsymbol{y}\in\mathbb{R}^N$ be the signal which is sparsely represented over the dictionary $\boldsymbol{D}\in\mathbb{R}^{N\times K}$ with $K>N$. This amounts to the following problem,
\begin{equation}\label{eq1}
\hat{\mathbf{x}}=\underset{\mathbf{x}}{\text{argmin}} \;\|\mathbf{x}\|_0 \; \;\text{s.t.} \; \;\mathbf{y}=\boldsymbol{D}\mathbf{x}
\end{equation}
where $\|.\|_0$  stands for the so-called $\ell_0$ pseudo-norm which counts the number of nonzero elements. Many algorithms have been introduced to solve the problem of finding the sparsest approximation of a signal in a given over-complete dictionary (for a good review see \cite{Tropp10}).
For a specified class of signals, e.g. class of natural images, the dictionary should have the capability of sparsely representing the signals. In some applications there is a predefined and fixed dictionary which is well-matched to the contents of the specific class of signals. Over-complete DCT dictionary for the class of natural images is an example. These non-adaptive dictionaries are favorable because of their simplicity. On the other hand, learning based dictionary results in better matching the contents of the signals \cite{Elad1}. Most dictionary learning algorithms are indeed a generalization of the clustering algorithms. While in clustering each training signal is forced to assign only one atom (cluster center), in the dictionary learning problem each signal is allowed to use more than one atom provided that it uses as fewest as few atoms as possible. The general dictionary learning problem can be stated as follows,
\begin{equation}\label{eq2}
(\hat{\boldsymbol{D}},\hat{\boldsymbol{X}})=\underset{\boldsymbol{D,X}}{\text{argmin}} \;\|\boldsymbol{Y}-\boldsymbol{D}\boldsymbol{X} \|_F^2 \; \;\text{s.t.} \; \;\|\mathbf{x}_i\|_0 \le T \; \forall i.
\end{equation}
Where the columns of $\boldsymbol{Y}$ contain the observed data and $\mathbf{x}_i$, the columns of $\boldsymbol{X}$, are sparse representations of the observed data. Most dictionary learning algorithms solve the above problem by alternatively minimizing it over $\boldsymbol{D}$ and $\boldsymbol{X}$. Dictionary learning algorithms differ mainly in performing the minimization over the dictionary. Dictionary learning has an important role in the sparse decomposition based methods. A subsection in the proposed method section is allocated for discussion on dictionary learning.

\subsection{Basic Theory of Detection}
\begin{figure}[b]
\centering
    \includegraphics[height=1.2in, width=6.5in]{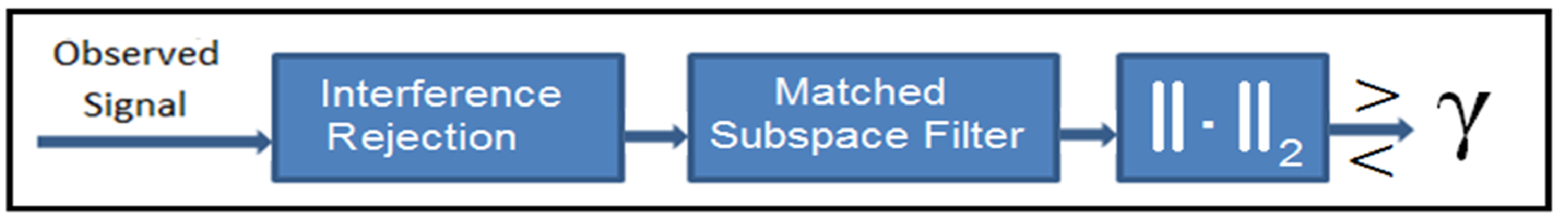}
  \caption{Block diagram of the matched subspace detector.}
  \label{matched_det}
\end{figure}

\begin{figure}
\centering
    \includegraphics[height=3.5in, width=7in]{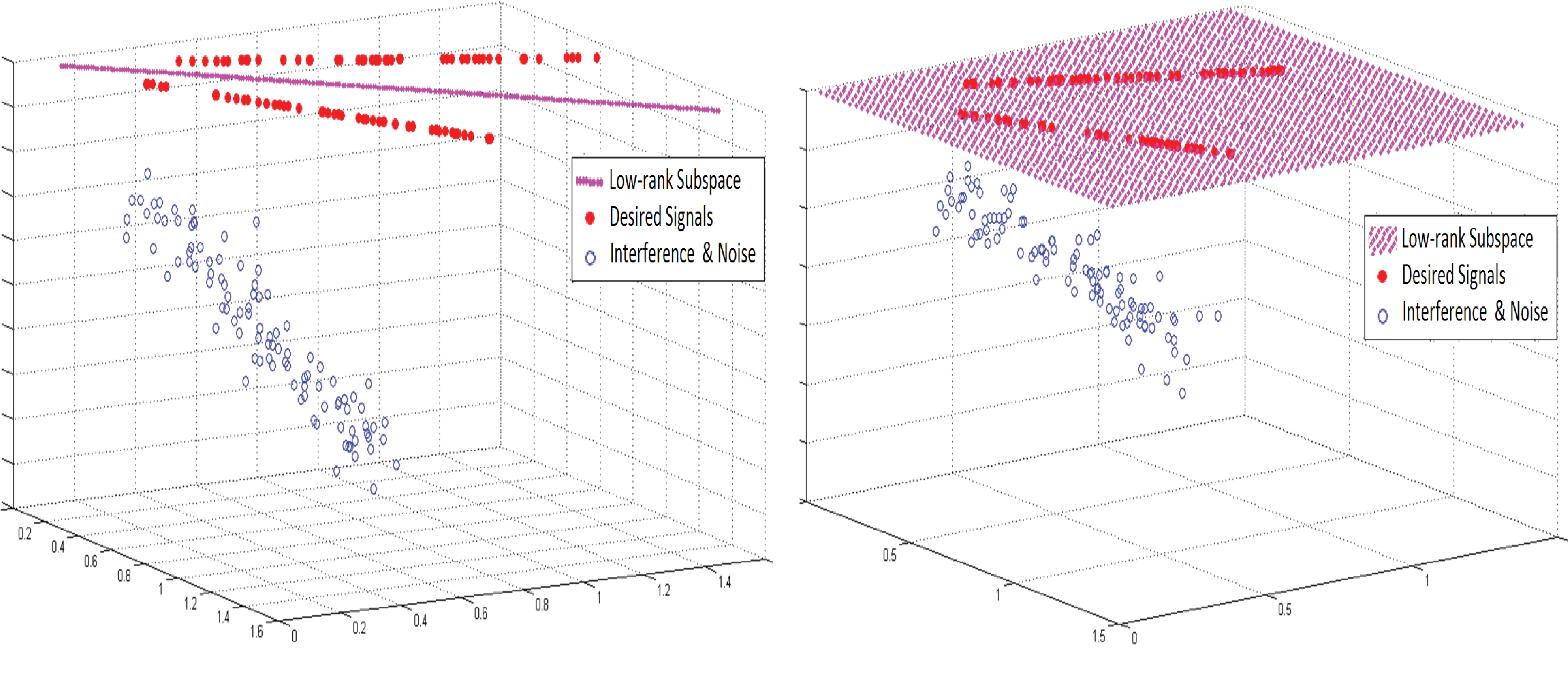}
  \caption{Two low-rank subspaces learned by PCA for some 3D signals.}
  \label{pca_sub}
\end{figure}

\begin{figure}[t]
\centering
    \includegraphics[height=3.5in, width=4in]{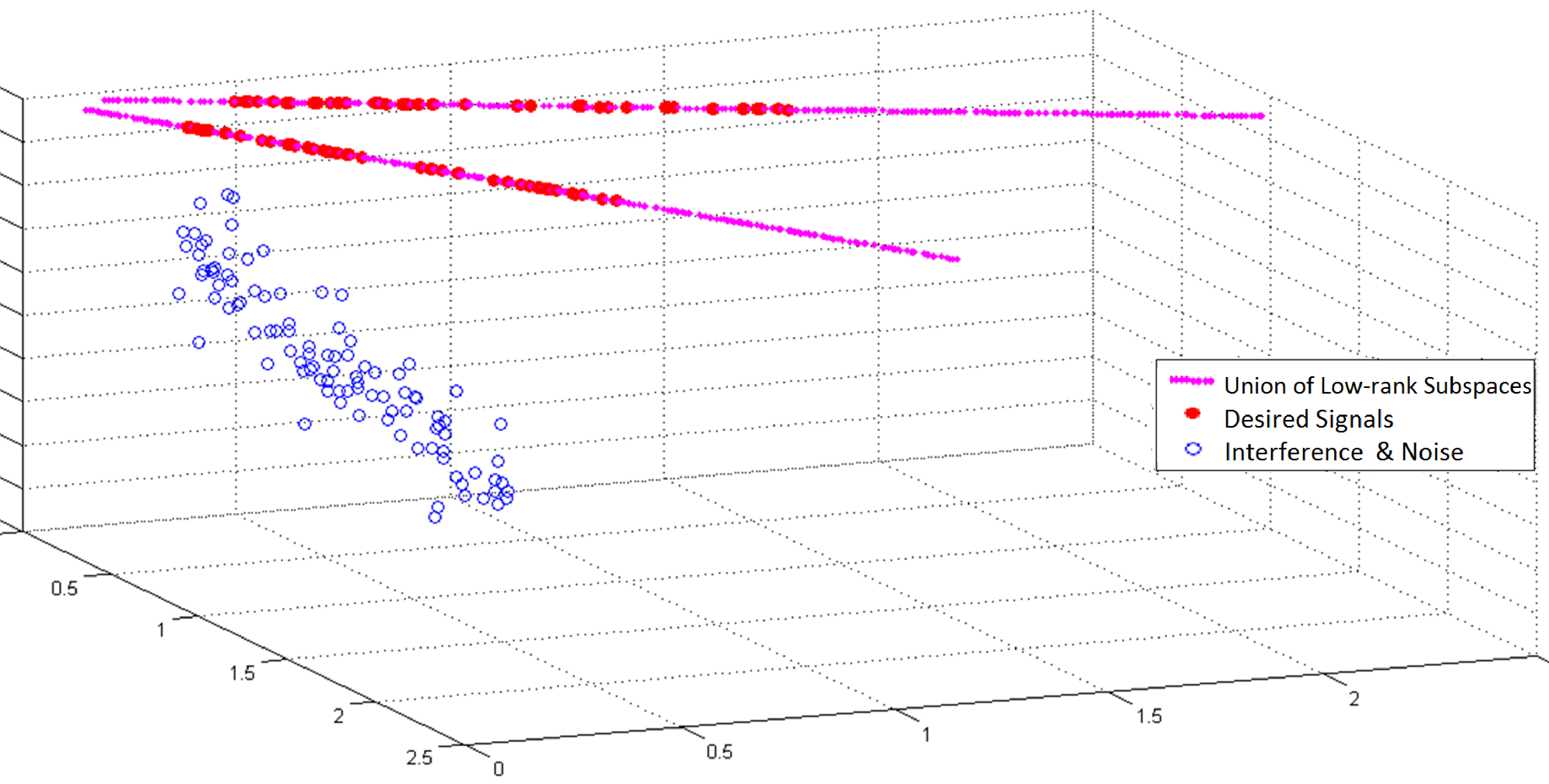}
  \caption{A union of rank-1 subspaces provides suitable matched subspaces.}
  \label{uls_sub}
\end{figure}

In this section we review signal detection theory and study some related detectors to our proposed one. First consider the following model for detection
\begin{equation}\label{eq3}
\begin{split}
&\mathfrak{H_0}: \mathbf{y}=\mathbf{n} \quad \quad \quad \quad \text{signal absence} \\
&\mathfrak{H_1}: \mathbf{y}=\mathbf{s}+\mathbf{n} \quad \quad \; \text{signal presence}
\end{split}
\end{equation}
where $\boldsymbol{y}\in\mathbb{R}^N$ is the observation vector, $\boldsymbol{s}\in\mathbb{R}^N$ is the signal of interest and $\boldsymbol{n}\in\mathbb{R}^N$ is the observation noise of the model. First we assume that the probability density function of $p(y|\mathfrak{H}_0)$ and $p(y|\mathfrak{H}_1)$ are known. In this case the likelihood ratio test (LRT) gives
\begin{equation}\label{eq4}
\frac{p(\mathbf{y}|\mathfrak{H_1})}{p(\mathbf{y}|\mathfrak{H_0})} \underset{\mathfrak{H_0}}{\overset{\mathfrak{H_1}} {\lessgtr}} \gamma
\end{equation}
where $\gamma$ is a threshold that satisfies the desired amount of probability of false alarm. By Gaussian assumption on the noise with covariance matrix R, LRT simplifies to,
\begin{equation}\label{eq5}
\mathbf{s}^H\boldsymbol{R}^{-1}\mathbf{y}\lessgtr \gamma
\end{equation}
where $\boldsymbol{R}$ is the covariance matrix. If it is not known in advance, it must be determined by obtaining the sample covariance matrix in the above test. Probability of detection is then equal to \cite{Davenport06},
\begin{equation}\label{eq6}
P_D(\alpha)=Q(Q^{-1}(\alpha)-\sqrt{SNR})
\end{equation}
in which $\alpha$ is the probability of false alarm and $Q(\alpha)=\int_\alpha^{\infty} \text{exp}(-t^2)\text{d}t$.
Another well-known detector is Generalized LRT \cite{Kelly11} (GLRT) which is derived by maximizing conditional densities constituting the likelihood ratio test with respect to the unknown parameters. The following detection criterion is obtained by assuming the covariance matrix to be unknown
\begin{equation}\label{eq7}
{|\mathbf{s}^H\boldsymbol{\hat{R}}^{-1}\mathbf{y}| \over \mathbf{s}^H\boldsymbol{\hat{R}}^{-1}\mathbf{s} (1+\frac{1}{K}\mathbf{y}^H\boldsymbol{\hat{R}}^{-1}\mathbf{y})}\lessgtr \gamma
\end{equation}
where $L$ is the number of snapshots available for $\hat{R}$ estimation. In \cite{Kelly11} no optimality has been claimed for GLRT. However, Scharf and Friedlander have shown that GLRT is uniformly most powerful (UMP) invariant \cite{Scharf12}. This is the strongest statement of optimality derived  for a detector. GLRT detector may be interpreted as a projection on the null-space of the interference followed by a matched subspace detector \cite{Scharf12}. Consider the following model for hypothesis test.
\begin{equation}\label{eq8}
\begin{split}
&\mathfrak{H_0}: \mathbf{y}=\boldsymbol{C\theta}+\mathbf{n} \quad \quad \quad \quad \quad \; \text{signal absence} \\
&\mathfrak{H_1}: \mathbf{y}=\boldsymbol{D}\mathbf{x}+\boldsymbol{C\theta}+\mathbf{n} \quad \quad \; \text{signal presence}
\end{split}
\end{equation}
where $\boldsymbol{C}\in\mathbb{R}^{N\times p}$ spans the background or interference subspace, $p<N$ and $\theta$ determines contribution of each column of $C$. $\boldsymbol{D}\in\mathbb{R}^{N\times m}$ spans signal subspace which is to be detected, $m<N$ and $\textbf{x}$ determines the contribution of each column of $\boldsymbol{D}$. It is obvious that if $p\ge N$ or $m\ge N$   then $\boldsymbol{C}$ or $\boldsymbol{D}$ spans all the space of the signals. In other words, in this case $\boldsymbol{C}$ or $\boldsymbol{D}$ may be over-fitted for background detection and signal detection, respectively. On the other hand, restriction of $p$ and $m$ may result in unreliable subspaces which are unable to fit suitable matched subspace. The role of matched subspaces detector is as follows
\begin{equation}\label{eq9}
\mathbf{y}^H P_C^{\perp} P_{DC^{\perp}}P_C^{\perp}\mathbf{y} \lessgtr \gamma
\end{equation}
where $P_C^{\perp}$ is the orthogonal projection matrix on the null-space of $\boldsymbol{C}$ and $P_{DC^{\perp}}$ is the part of orthogonal projection of $P_D$ which does not account for subspace spanned by $\boldsymbol{C}$. Figure \ref{matched_det} shows the block diagram of this detector.
At the conclusion of paper \cite{Scharf12} authors mentioned that basis can be extracted from Discrete Cosine Transform, Wavelet Transform or learned by data dependent analysis like Principal Component Analysis (PCA). Using such basis provides a matched subspace for the whole desired signals which are going to be detected. For more illustration refer to Figure \ref{pca_sub}. This figure shows composites of some 3D data by signal and non-signal (interference and noise) parts. Two low-rank subspaces are shown corresponding to rank-1 and rank-2 subspace (the low-rank matched subspace) which are obtained by PCA.
The main contribution of \cite{Scharf12} may be answering 'no' to the question, 'Can the GLRT be improved upon?' while they did not assume any prior information on the structure of the low-rank matched filter. The structural assumption can be applied by assuming a sparse prior on the coefficients of $\boldsymbol{\theta}$ and $\boldsymbol{x}$. The proposed method of this paper suggests using the model of ULRS for signals due to its suitable fitness which has been proven in many signal processing applications. Instead of traditional analysis like PCA, modern analysis like the methods proposed in \cite{Sadeghi11755}, \cite{Aharon13} and \cite{Rubinstein14} can be exploited in order to recover suitable bases spanning these low-rank subspaces. Figure \ref{uls_sub} shows a union of matched low-rank subspaces corresponding to the data of Figure \ref{pca_sub}.

Compressive detection is another application of the sparse theory exploited in signal detection and studied in \cite{Wang15} and \cite{Davenport16}. Instead of dealing with all the samples of the signal, the compressed detector works with few measurements. This detector distinguishes between two hypotheses,
\begin{equation}\label{eq10}
\begin{split}
&\mathfrak{H_0}: \mathbf{z}=\boldsymbol{\Phi}\mathbf{n} \quad \quad \quad \quad \; \text{signal absence} \\
&\mathfrak{H_1}: \mathbf{z}=\boldsymbol{\Phi}(\mathbf{s}+\mathbf{n}) \quad \quad \text{signal presence}
\end{split}
\end{equation}
where $\boldsymbol{\Phi} \in \mathbb{R}^{M\times N}$ is the measurement matrix and $\boldsymbol{z}$ is the measurement. If no further prior is known about $\boldsymbol{s}$, no optimal $\boldsymbol{\Phi}$ can be designed, and random measurements yield a detector with the following performance \cite{Wang15}.
\begin{equation}\label{eq11}
P_D(\alpha)\cong Q(Q^{-1}(\alpha)-\sqrt{\frac{M}{N}SNR})
\end{equation}
in which, the performance of the detector is degraded by factor $\sqrt{\frac{M}{N}}$ compared to the traditional matched filter. Having knowledge of $\boldsymbol{s}=\boldsymbol{D}\boldsymbol{\theta}$ results in a compressed detector as shown in \cite{Wang15}.
\begin{equation}\label{eq12}
P_D(\alpha)\cong Q(Q^{-1}(\alpha)-\sqrt{\frac{M}{K}SNR})
\end{equation}

in which, the performance of the detector is improved by a factor of $\sqrt{\frac{M}{N}}$ compared to the random measurement detector. Reference \cite{Davenport16} studied two cases about the knowledge of $\boldsymbol{D}$. The first case assumes that $\boldsymbol{D}$ is known and the second case assumes that $\boldsymbol{D}$ consists of a set of parametric basis, where the active basis of $\boldsymbol{D}$ can be recovered by a sparse coding algorithm. Recently, \cite{Eldar17} investigated the problem of detection of a union of low-rank subspaces via compressed measurements. The compressed detector still performs worse than the matched filter by factor $\sqrt{\frac{M}{N}}$.

In this paper we are going to exploit the low-rank structure characteristic of the signals to design a new detector. Our detector is not compressed and the goal is to design a generalized detector using sparsity (that is, assuming a structure) which implicitly exists in the signals. In Section 3 the proposed  detector will be presented. Our detector first assumes a model according to sparse signals and then derives an optimum rule of detection.
\section{The Proposed Approach}
In this section, we introduce our model for signal detection. We want to distinguish between two hypotheses $\mathfrak{H}_0$ and $\mathfrak{H}_1$:
\begin{equation}\label{eq13}
\begin{split}
&\mathfrak{H_0}: \mathbf{y}=\mathbf{n} \quad \quad \quad \quad \quad\quad\quad \; \text{signal absence} \\
&\mathfrak{H_1}: \mathbf{y}=\boldsymbol{D}\mathbf{x}+\mathbf{e}+\mathbf{n} \quad \quad \; \;\text{signal presence}
\end{split}
\end{equation}

where, $\boldsymbol{D}$ is the dictionary which can be interpreted as a bank of matched filters, $\mathbf{e}$ is the error vector of the model which denotes the mismatch between the exact matched filter and the union of subspaces spanned by the columns of $\boldsymbol{D}$. Assume that $\mathbf{e}$ is a zero-mean white Gaussian noise with variance $\sigma_e^2$ , i.e. $\mathbf{e}~N(0,\sigma_e^2 \boldsymbol{I})$. In our method, the signal ($\boldsymbol{D}\mathbf{x}$) matched to the observed signal ($\mathbf{y}$) is unknown; so it must be determined.
This section is divided into four subsections.  In the first subsection, we analyze the role of the coefficients of the linear combination ($\mathbf{x}$) and then describe our approach for coefficients estimation. In the second subsection, the performance of our proposed detection method will be analyzed. Since dictionary learning is a critical issue in the model, third subsection is allocated for discussing on the dictionary learning. In the last subsection we will explain how our method may become robust to detect signals that are contaminated by gross errors.
\subsection{A discussion on the coefficients ($\mathbf{x}$) }
Linear combination of the dictionary atoms generates the matched signal for detection. Three cases are considered for $\mathbf{x}$ estimation. First, no constraint solution, second matched filter bank and third applying Gaussian distribution. First assume that there is no constraint on $\mathbf{x}$ i.e, orthogonal projection of the signal onto the span of the desired subspace. This method is used in matched subspace method to identify the part of signal that amounts for the desired signals \cite{Scharf12}. The solution for it will be,
\begin{equation}\label{eq14}
\hat{\mathbf{x}}=\underset{\mathbf{x}}{\text{argmin}}  \|\mathbf{y}-\boldsymbol{D}\mathbf{x}\|_2^2 = (\boldsymbol{D}^T\boldsymbol{D})^{-1} \boldsymbol{D}^T\mathbf{y}
\end{equation}
This answer suffers from over-fitting as some signals that do not contain the target signal may be decomposed in terms of the atoms. More restricted constraints may alleviate this problem. Now let us assume that just one element of $\mathbf{x}$ is allowed to be none zero. This constraint helps reducing over-fitting. By this assumption the problem becomes,
\begin{equation}\label{eq15}
\hat{\mathbf{x}}=\underset{\mathbf{x}}{\text{argmin}}  \|\mathbf{y}-\boldsymbol{D}\mathbf{x}\|_2^2 \quad s.t. \quad \|\mathbf{x}\|_0=1
\end{equation}
The solution will be zero except in the position corresponding to the atom with maximum correlation. This solution is nothing but the traditional matched filter bank. Each matched filter which has more correlation is considered as the matched signal. All correlations are sufficient statistics for the decision. If all the correlations are less than a threshold, no detection is performed.

The third scenario we study is assuming Gaussian prior on $\mathbf{x}$. The motivation of considering this assumption for $\mathbf{x}$ is to avoid over-learning and moreover having less sensitive coefficients. Estimation of $\mathbf{x}$ by the assumption of Gaussian distribution on $\mathbf{n}$ and $\mathbf{x}$ can be obtained as follows,
\begin{equation}\label{eq16}
\hat{\mathbf{x}}=\underset{\mathbf{x}}{\text{argmin}}  \|\mathbf{y}-\boldsymbol{D}\mathbf{x}\|_2^2 +\lambda \|\mathbf{x}\|_2^2 = (\boldsymbol{D}^T\boldsymbol{D}+\lambda \boldsymbol{I})^{-1} \boldsymbol{D}^T\mathbf{y}
\end{equation}
This solution for the coefficients of linear combination is the Ridge regression \cite{Hoerl18}. Solution (\ref{eq15}) is the least over-learned and solution (\ref{eq14}) is the most over-learned one. It is interesting to see how each of the solutions covers the signal space for learning. Solution (\ref{eq15}) provides high learning for few one dimensional subspaces corresponding to each atom, while solutions (\ref{eq14}) and (\ref{eq16}) provide high learning for many subspaces corresponding to arbitrary selections of the atoms. Involvement of all the atoms to form the matched signal results in detection of undesired signals as the target signal due to the expansion of the matched subspaces. To keep the number of involved atoms limited, we suggest modifying problem (\ref{eq16}) as follows,
\begin{equation}\label{eq17}
\hat{\mathbf{x}}=\underset{\mathbf{x}}{\text{argmin}}  \|\mathbf{y}-\boldsymbol{D}\mathbf{x}\|_2^2 +\lambda \|\mathbf{x}\|_0
\end{equation}
There is a large enough value for $\lambda$ such that the solution of the above problem is the same as (15). Now we show that this problem is the MAP estimation of $\mathbf{x}$ under multivariate independent Gaussian prior,
\begin{equation}\label{eq18}
p(\mathbf{x},\boldsymbol{W})\propto \sqrt{|\boldsymbol{W}|}\; \text{exp}(-\mathbf{x}^T\boldsymbol{W}\mathbf{x})
\end{equation}
where $\boldsymbol{W}$ is a diagonal matrix. By this assumption, two unknowns must be estimated. First we obtain the ML estimation of $\boldsymbol{W}$,
\begin{equation}\label{eq19}
\boldsymbol{W}_{ML}=\underset{\boldsymbol{W}}{\text{argmax}} \; \text{log}\;p(\mathbf{y}|\mathbf{x},\boldsymbol{W})=\underset{\boldsymbol{W}}{\text{argmax}}\; \mathbf{x}^T\boldsymbol{W} \mathbf{x} -\text{log}(|\boldsymbol{W}|)
\end{equation}
By setting the derivative with respect to $\boldsymbol{W}$ equal to zero, the solution of Equation (\ref{eq19}) is $\boldsymbol{W}^{-1}=\mathbf{x}\mathbf{x}^T$ which has no solution, however we need only the diagonal elements of $\boldsymbol{W}$ due to the independent assumption on the entries of $\mathbf{x}$. So, calculating the derivative with respect to only diagonal elements of $\boldsymbol{W}$ ($w_{ii}$) results in,
\begin{equation}\label{eq20}
w_{ii}=\frac{1}{x_i^2}=\lim_{\delta\rightarrow 0} \frac{1}{\delta + x_i^2}
\end{equation}
where, $\delta$ is a small positive for avoiding division by zero. Then we insert the obtained $\boldsymbol{W}$ in (18):
\begin{equation}\label{eq21}
p(\mathbf{x})\propto \text{exp} (-\sum \lim_{\delta\rightarrow 0} \frac{x_i^2}{\delta + x_i^2})=\text{exp} (-\|\mathbf{x}\|_0)
\end{equation}

\begin{figure}
\centering
    \includegraphics[height=2.2in, width=4.5in]{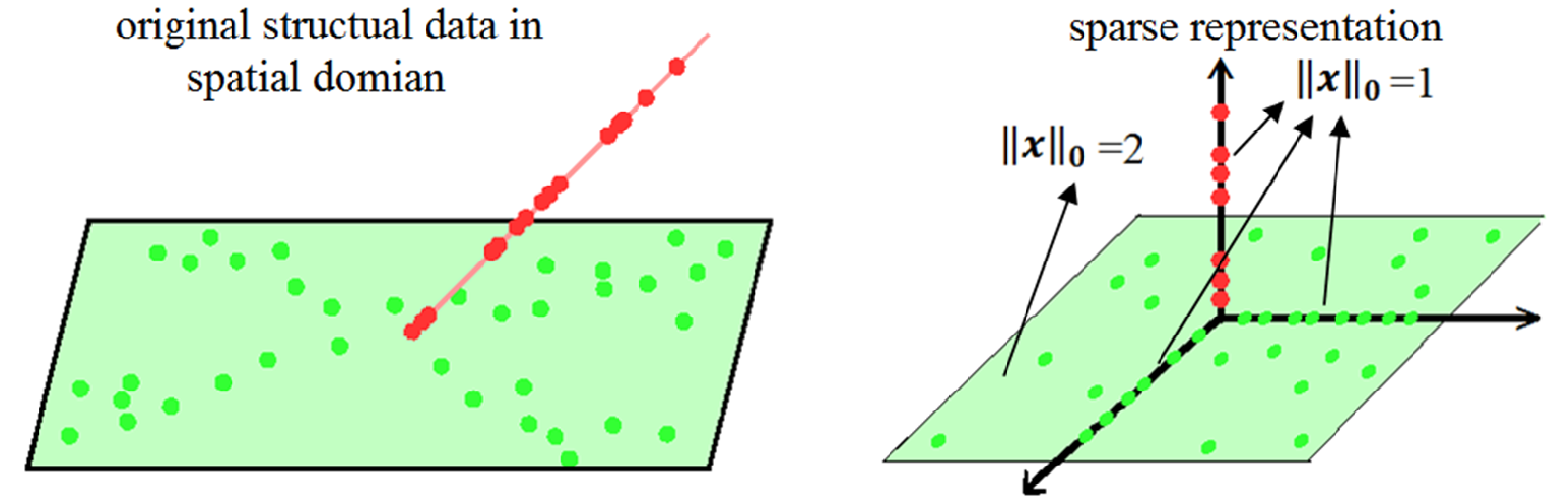}
  \caption{Sparse representation of some structural data whose distribution is in agreement with the one defined by Eq. (\ref{eq21}).}
  \label{subspaces}
\end{figure}
Actually, $\boldsymbol{W}$ is an auxiliary parameter which is used just for more adaptation of the coefficients distribution. The obtained W results in a distribution with more probability of having orthogonal low rank subspaces (in the space to which $\mathbf{x}$ belongs; for more illustration see Fig.  4). Corresponding to these orthogonal low rank subspaces there are non-orthogonal low rank subspaces in the observation domain which $\mathbf{y}$ or $\boldsymbol{D}\mathbf{x}$ belongs to this space.
The MAP estimation of $\mathbf{x}$ by prior of (\ref{eq21}) results in the suggested problem (\ref{eq17}) which is a generalized version of (\ref{eq15}) from the aspect of sparsity level of the coefficients, and a generalized version of (\ref{eq16}) from the aspect of prior distribution on the coefficients for estimation. In \cite{Elad4}, it is proved that in a certain condition, problem (\ref{eq17}) leads to the same solution with the following $\ell_1$ regularized  problem:
\begin{equation}\label{eq22}
\hat{\mathbf{x}}=\underset{\mathbf{x}}{\text{argmin}}  \|\mathbf{y}-\boldsymbol{D}\mathbf{x}\|_2^2 +\lambda \|\mathbf{x}\|_1
\end{equation}
\subsection{Performance analysis}
First, we define the false alarm rate and the detection alarm rate,
\begin{equation}\label{eq23}
\begin{split}
P_F=\text{Pr}(\mathfrak{H}_1 \text{is chosen while } \mathfrak{H}_0 \text{ is true})\\
P_D=\text{Pr}(\mathfrak{H}_1 \text{is chosen while } \mathfrak{H}_1 \text{ is true})\\
P_F=\int_{p(\mathbf{y|\mathfrak{H}_1})>\gamma p(\mathbf{y|\mathfrak{H}_0})} p(\mathbf{y|\mathfrak{H}_0}) \text{d}\mathbf{y} = \alpha.
\end{split}
\end{equation}
Parameter $\gamma$ satisfies the desired amount of false alarm probability, $\alpha$.
\begin{equation}\label{eq24}
\begin{split}
&p(\mathbf{y}|\mathfrak{H}_0)=\frac{1}{(2\pi \sigma_n^2)^{N/2}}\text{exp} (-\frac{\|\mathbf{y}\|_2^2}{2\sigma_n^2})\\
&p(\mathbf{y}|\mathfrak{H}_1)=\frac{1}{(2\pi (\sigma_n^2 + \sigma_e^2))^{N/2}}\text{exp} (-\frac{\|\mathbf{y}-\boldsymbol{D}\mathbf{x}\|_2^2}{2(\sigma_n^2 +\sigma_e^2)})
\end{split}
\end{equation}
By solving $p(y|\mathfrak{H}_1)=\gamma p(y|\mathfrak{H}_0 )$, the threshold for decision rule can be achieved,
\begin{equation}\label{eq25}
t=<\mathbf{y},\boldsymbol{D}\mathbf{x}> \underset{\mathfrak{H_0}}{\overset{\mathfrak{H_1}} {\lessgtr}} C+\|\boldsymbol{D}\mathbf{x}\|_2^2-\|\mathbf{y}\|_2^2(\frac{\sigma_e^2}{\sigma_n^2}),
\end{equation}
where $C$ is a constant value depending on $\sigma_n^2$ and $\sigma_e^2$ and the desired $\alpha$. The sufficient statistic for decision making is $t=<\boldsymbol{y},\boldsymbol{Dx}>=\boldsymbol{y}^T\boldsymbol{Dx}$.  It is easy to show that,
\begin{equation}\label{eq26}
P_D(\alpha)=Q(Q^{-1}(\alpha)-\sqrt{\frac{\text{SNR}}{1+\text{ESR}}})
\end{equation}
where $\text{SNR}=\mathbb{E}{\|\boldsymbol{D}\mathbf{x}\|_2^2}/\sigma_n^2 $ and $\text{ESR}=\sigma_e^2/\mathbb{E}{\|\boldsymbol{D}\mathbf{x}\|_2^2}$. As can be seen, the performance of the detector is degraded by a factor of $1+\text{ESR}$. But our detector has learned a suitable space for signals to be detected. In other words, we accept a small deterioration of the performance duo to the generalization of the detector. Flexibility of the sparse representation based detector is the most distinguished advantage. Dictionary learning \cite{Aharon13} is the most important issue for the methods based on sparse representation. In the sparse detector, the dictionary should be learned such that ESR$\ll 1$ to avoid performance deterioration and at the same time ESR$\gg \epsilon$ to avoid over-learning. In the next section we will explain how to learn an appropriate dictionary. In (25), sparsity has no effect on the performance. Now we introduce a decision rule for detection that exploits the sparsity of the coefficients. To this end, we solve equation $p(\mathbf{y}|\mathbf{x},\mathfrak{H}_1)p(\mathbf{x})=\gamma p(\mathbf{y}|\mathfrak{H}_0 )$ by the obtained $p(\mathbf{x})$ in the equation (21). The new decision rule can be achieved as follows,
\begin{equation}\label{eq27}
t=<\mathbf{y},\boldsymbol{D}\mathbf{x}> \underset{\mathfrak{H_0}}{\overset{\mathfrak{H_1}} {\lessgtr}} C+\|\boldsymbol{D}\mathbf{x}\|_2^2-\|\mathbf{y}\|_2^2(\frac{\sigma_e^2}{\sigma_n^2}) +\gamma \|\mathbf{x}\|_0
\end{equation}
where $c$ is a positive constant value. As $\|\mathbf{x}\|_0$ increases, $\mathfrak{H}_0$ may be more probable, because the signals representation in terms of the dictionary would be sparse only for the learned signals. Similar to (\ref{eq26}), it is easy to show that,
\begin{equation}\label{eq28}
P_D(\alpha)=Q(f(\|\mathbf{x}\|_0) Q^{-1}( \alpha)-\sqrt{\frac{\text{SNR}}{1+\text{ESR}}})
\end{equation}
where $f$ is an increasing homogenous function. As can be seen, the probability of detection increases (decreases) when sparsity increases  (decreases) for false alarm rates smaller than  $0.5$ (because when $\alpha<0.5$ then $Q^{-1} (\alpha)>0)$). As the desired false alarm rates are often small, the probability of detection would increase in this region (it is favorable for a detector that the top-left region of its ROC be close to the ideal ROC). If the representation of a signal is sparse, this signal lies in the desired low-rank subspace (that is, it meets our assumed model for the target signals). Thus the probability of detection would increase for these signals that have sparse representation in terms of the dictionary atoms, which is actually what we expect from sparsity. Figure 5 shows the ROC of (\ref{eq22}) with SNR=+20dB for different sparsity levels.
\begin{figure}
\centering
    \includegraphics[height=4in, width=6in]{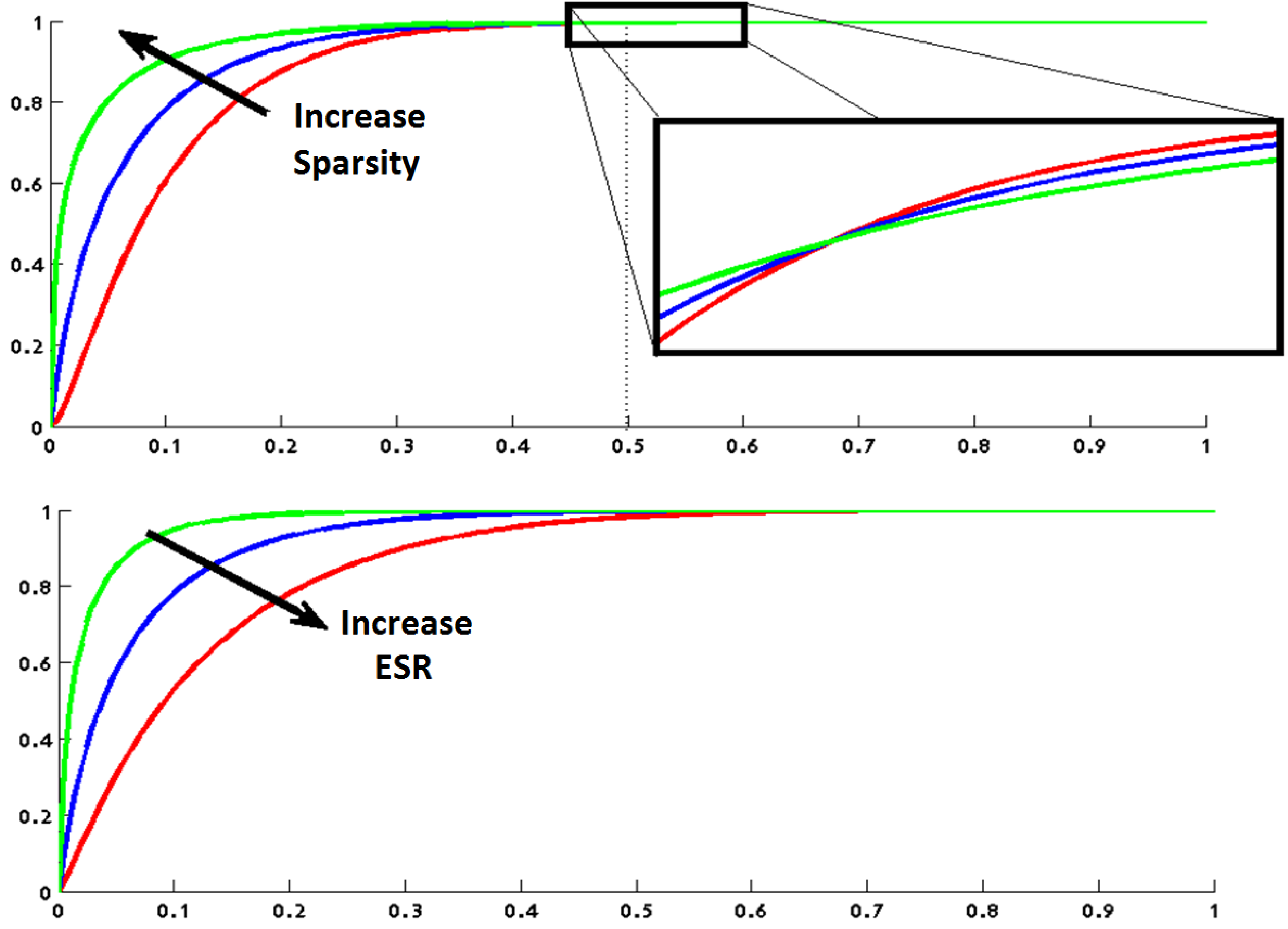}
  \caption{Trade-off between sparsity and ESR and its effect on the detector performance.}
  \label{ROC1}
\end{figure}
Traditional matched filter banks have the most sparsity level, but it is not practical. For instance, in voice activity detection, it is not feasible to collect all possible voices in a bank. A small number of filters results in high ESR and low performance. Our proposed detector makes a trade-off between ESR and sparsity in order to have a good detector performance. Dictionary learning has a critical role in the trade-off which is studied in the following.
\subsection{Learning the Dictionary}
In this section we explain the role of dictionary learning in the proposed detection method. In many detection problems, the number of training signals may not be as large as the number of possible matched filters that cover all the target signals space. By the proposed approach, we search for a dictionary learned by a set of finite number of signals that efficiently represents those signals. The dictionary should be general to be able to deal with a signal that has not been seen before. Assume that we have a set of signals ($\boldsymbol{Y}\in \mathbb{R}^{N\times L}$). Dictionary learning is a function that maps  $\boldsymbol{Y}$ to $\boldsymbol{D}$ where $\boldsymbol{D}\in \mathbb{R}^{N\times K}$. An appropriate dictionary should have ESR$\ll 1$ to be a suitable representation for the training data and also ESR should not be too small to have a general dictionary that is not over-learned for only the training data. Two algorithms for dictionary learning are presented.
\subsubsection{K-means algorithm}
K-means method uses K centroids of clusters, to characterize the training data \cite{Jain19}. They are determined by minimizing the sum of squared errors,
\begin{equation}\label{eq29}
\boldsymbol{D}=\underset{\boldsymbol{D}}{\text{argmin}}\; \sum_{k=1}^K \sum_{i\in C_k} \; \|\mathbf{y}_i -\mathbf{d}_k\|_2^2
\end{equation}
where the columns of $\boldsymbol{D}$ are $\boldsymbol{d}_k$, $k=1,\ldots,K$. The provided dictionary assigns to each training data a centroid. $K$ should be large enough to satisfy the desired amount of ESR. Problem (\ref{eq14}) has to be solved to determine the coefficients so that only one of them is none zero. This dictionary learns some points in the signal space. As the distance from these points increases, the level of learning would decrease. In other words, this dictionary is obtained by the union of spheres model. This model may not be a suitable choice for ordinary signals. The next algorithm agrees with a more appropriate model for the data. The KSVD learns the signal space with a union of low-rank subspaces.
\subsubsection{K-SVD algorithm}
\begin{figure}
\centering
    \includegraphics[height=2.5in, width=3.2in]{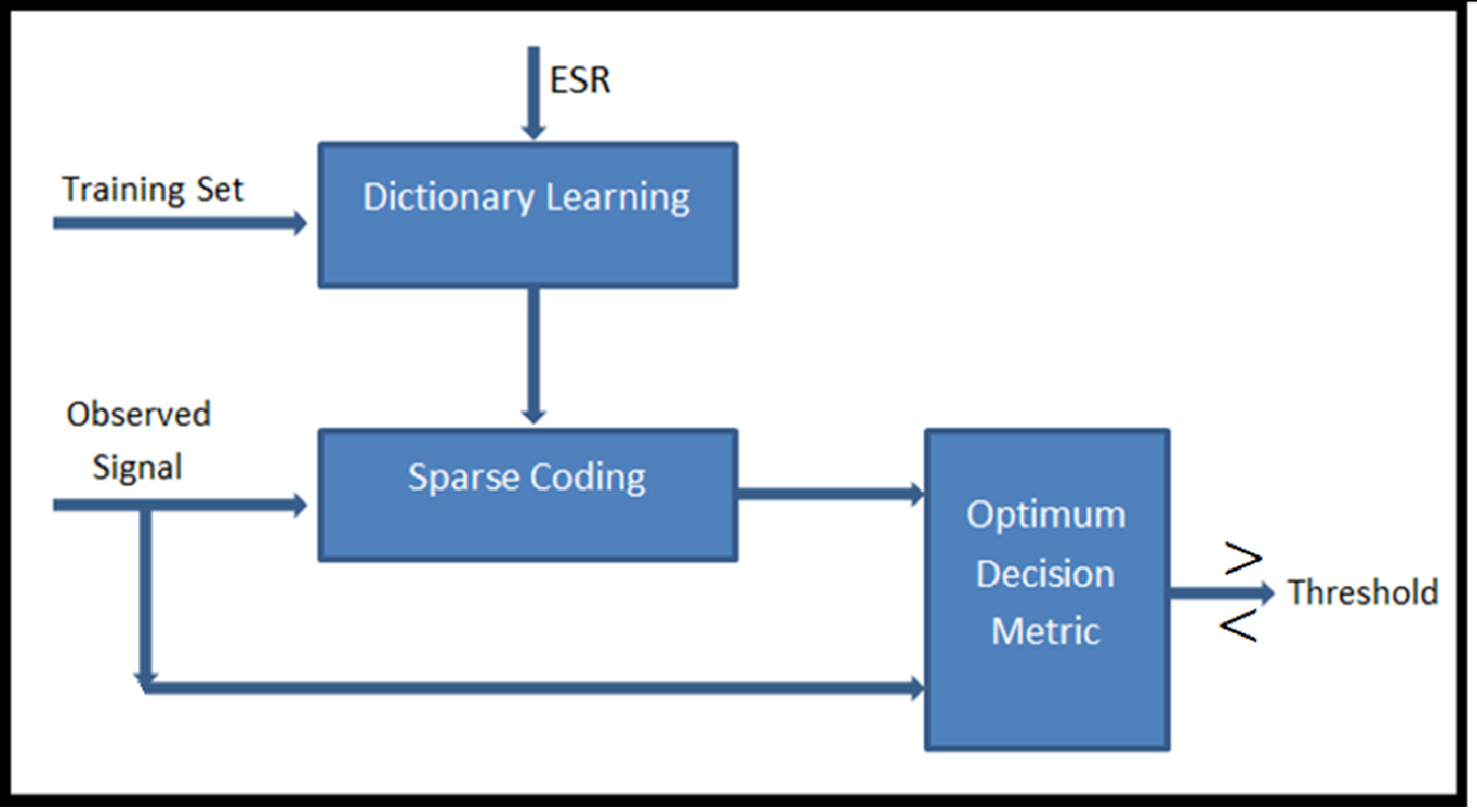}
  \caption{Block diagram of our proposed detector based on dictionary learning.}
  \label{diagram}
\end{figure}

By extending the union of spheres to a union of low-dimensional subspaces, K-means algorithm is generalized to K-SVD algorithm \cite{Aharon13}. This flexible model agrees with many signals such as images and audio signals. For example, natural images have sparse representation in terms of DCT dictionary. In other words, by combination of only a few DCT bases, it is possible to approximate the blocks of an image. The following problem provides the dictionary learned by K-SVD,
\begin{equation}\label{eq30}
\boldsymbol{D}=\underset{\boldsymbol{D},\mathbf{x}_i}{\text{argmin}}\; \sum_{i} \; \|\mathbf{y}_i -\boldsymbol{D} \mathbf{x}_i\|_2^2 \; \; \text{s.t.} \; \; \|\mathbf{x}_i\|_0\le T.
\end{equation}
This algorithm is based on atom-by-atom updating over the columns of $\boldsymbol{D}$. Recently, more efficient algorithms for atom-by-atom updating are suggested in \cite{Sadeghi_dic83}. Each arbitrary selection of few $\boldsymbol{D}$ columns characterizes a cluster corresponding to a subspace. The dictionary learned by K-SVD is in agreement with the proposed problem (\ref{eq22}). After learning, test signals that lie on the learned low dimensional subspaces can be reconstructed and detected.
In addition to dictionary learning using training signals, it is possible to design a dictionary using parametric functions \cite{Yaghoobi20}. Kernels of FFT and DCT are two examples from this class of dictionaries where bases sweep the parameter of frequency. Figure \ref{diagram} shows the block diagram of the proposed detection method.

\subsection{Robustness}
Assume that a dictionary has learnt to detect face images without sun glasses. If a face image with sun glasses is given to it for detection, gross error in the region of eyes may result in a wrong detection. To solve this problem, a distribution has to be supposed that has longer tail than Gaussian. Laplace distribution is our suggestion for the error distribution. Thus $\mathfrak{H}_1$ implies that the observed signal is the combination of few atoms of $\boldsymbol{D}$, a Laplace distributed error and a Gaussian distributed noise.
\begin{equation}\label{eq31}
\mathfrak{H}_1 \; : \quad \mathbf{y}=\boldsymbol{D}\mathbf{x}+\mathbf{e}+\mathbf{n}
\end{equation}
The problem of coefficients estimation for (\ref{eq22}) by new prior assumption has been already presented in robust statistics \cite{Huber21}.
\begin{equation}\label{eq32}
\hat{\mathbf{x}}=\underset{\mathbf{x}}{\text{argmin}}  \|\mathbf{y}-\boldsymbol{D}\mathbf{x}\|_H +\lambda \|\mathbf{x}\|_1
\end{equation}
where,
\begin{equation}\label{eq33}
\begin{split}
\|\mathbf{s}\|_H=\sum h(s_i)\\
h(s)=\left\{
	\begin{array}{ll}
		s^2  & \mbox{if } |s| \leq \rho \\
		\rho s & \mbox{if } |s|>\rho
	\end{array}
\right.
\end{split}
\end{equation}
In other words, small errors and large errors are penalized by $\ell_2$ norm and $\ell_1$ norm, respectively. $\lambda$ is the parameter of the mixture distribution of Gaussian and Laplace. Let re-write (\ref{eq32}) as follows,
\begin{equation}\label{eq34}
\hat{\mathbf{x}}=\underset{\mathbf{x}}{\text{argmin}}  \|\mathbf{y}-\boldsymbol{D}\mathbf{x} - \mathbf{e}\|_2^2 +\lambda \|\mathbf{x}\|_1 +\rho \|\mathbf{e}\|_1
\end{equation}
$$
\hat{\mathbf{x}}=\underset{\mathbf{x}}{\text{argmin}}  \|\mathbf{y}-[\boldsymbol{D} \; \boldsymbol{I}][
\begin{array}{ll}
\mathbf{x} \\
\mathbf{e}
\end{array}
]\|_2^2 +\lambda (\|\textbf{x}\|_1 + \|\frac{\rho}{\lambda}\textbf{e}\|_1)
$$
Let us define $\tilde{\textbf{e}}$ as $\frac{\rho}{\lambda}\mathbf{e}$.
\begin{equation}\label{eq35}
\hat{\mathbf{x}}=\underset{\mathbf{x}}{\text{argmin}}  \|\mathbf{y}-[\boldsymbol{D} \; \frac{\lambda}{\rho}\boldsymbol{I}][
\begin{array}{ll}
\mathbf{x} \\
\tilde{\textbf{e}}
\end{array}
]\|_2^2 +\lambda (\|\mathbf{x}\|_1 + \|\tilde{\textbf{e}}\|_1)
\end{equation}
By substitution of $\boldsymbol{B}=[\boldsymbol{D}\quad \frac{\lambda}{\rho} I]$ and $\mathbf{z}=[\mathbf{x}^T \; \tilde{\textbf{e}}^T]^T$, we have,
\begin{equation}\label{eq36}
\hat{\mathbf{z}}=\underset{\mathbf{z}}{\text{argmin}}  \|\mathbf{y}-\boldsymbol{B}\mathbf{z}\|_2^2 +\lambda \|\mathbf{z}\|_1
\end{equation}
This problem is similar to (\ref{eq22}) except that its dictionary is extended by scaled identity matrix. Identity matrix projects inappropriate parts of the signals onto corresponding coefficients. Inappropriate parts of the signals may be large errors or out of the desired subspace interferences or outlier data. Authors of \cite{Wright22} also intuitively have used the same dictionary to obtain a robust framework for face recognition. A same procedure can be pursued to learn robust dictionary by a set of unreliable data \cite{Amini58854}.

\section{Experimental Results}
We evaluated the performance of our proposed method in the case study of VAD. To construct the learned dictionary, clean speech signals of NOIZEUS database were used \cite{Loizou23}. In NOIZEUS database, thirty sentences were selected which include all phonemes in the American English language. The sentences were produced by three male and three female speakers and originally sampled at 25 kHz and down-sampled to 8 kHz. We divided the clean speech signals into 25-ms frames with 10-ms frame shift. After removing the silent frames, we extracted standard Mel-frequency Cepstral Coefficients (MFCC) using 10 Mel triangular filters, energy values computed at each of the 10 Mel triangular filters, total energy (the first Cepstral coefficient) and entropy from each speech frame. MFCC features capture the most relevant information of speech signal, and they are widely used in speech and speaker recognition making the VAD method easy to integrate with existing applications. So our features vector was 24-dimensional, and the total number of vectors was about 6300. By using the K-SVD algorithm, we obtained a learned dictionary with 100 atoms, which was used in the following experiments for obtaining the sparse representation based on OMP method.
\begin{figure}[h]
\centering
    \includegraphics[height=3.7in, width=6.4in]{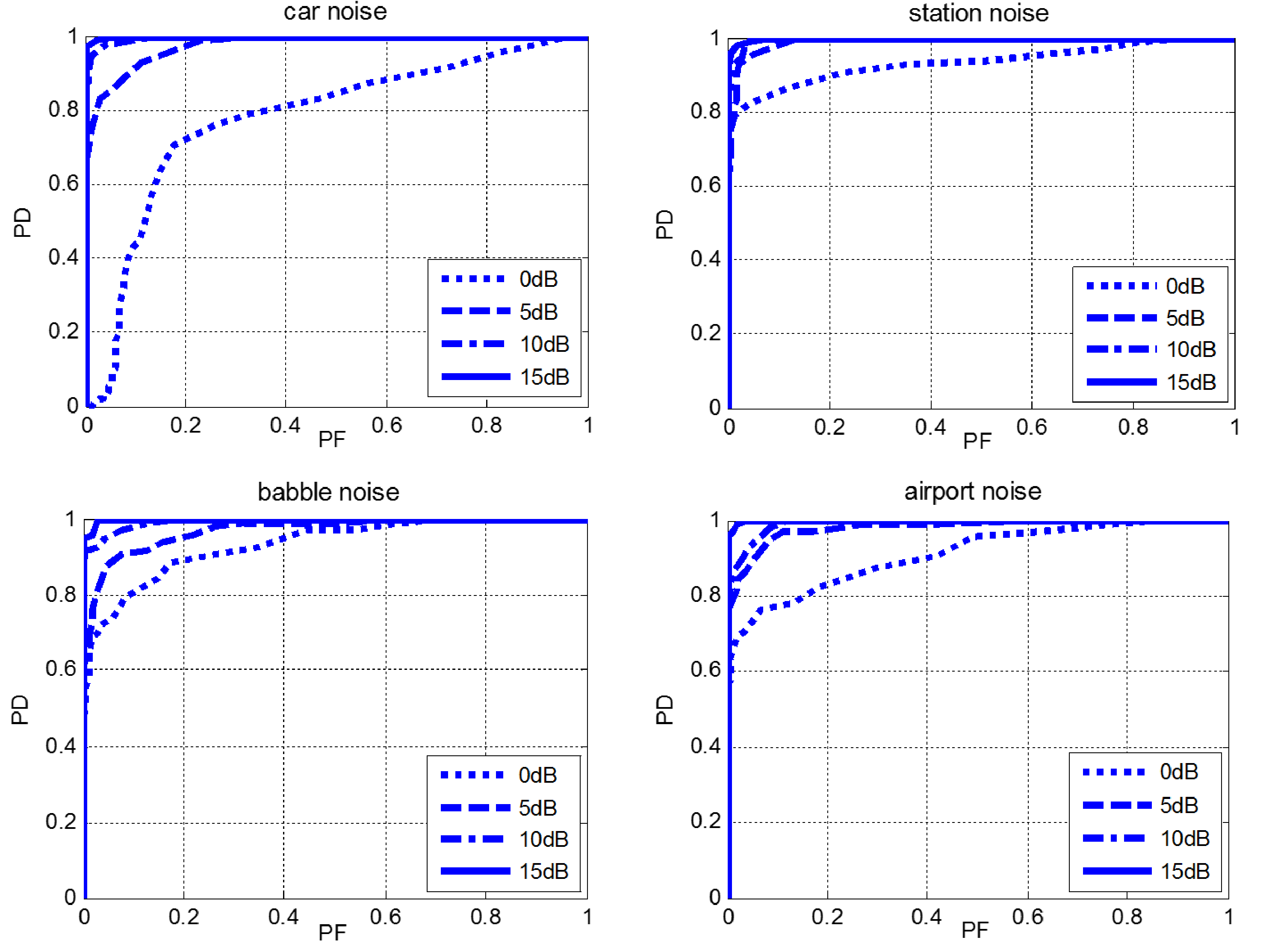}
  \caption{Block diagram of our proposed detector based on dictionary learning.}
  \label{simul1}
\end{figure}
To evaluate the performance of the proposed method, the speech detection probability PD and false alarm probability PF were investigated based on a reference decision. A clean test speech (sp10.wav), taken from the NOIZEUS database, was down-sampled at 8000 Hz and was used for the reference decisions. To simulate noisy environments, several noise signals as the subset of the NOIZEUS database were used. Noise signals included recordings from different places (Babble (crowd of people), Car,…) at SNRs of 0dB, 5dB, 10dB, and 15dB. The ROC Curves for VAD using our proposed method are illustrated in Fig. \ref{simul1} which shows PD versus PF.

\begin{figure}[h]
\centering
    \includegraphics[height=3.7in, width=6.6in]{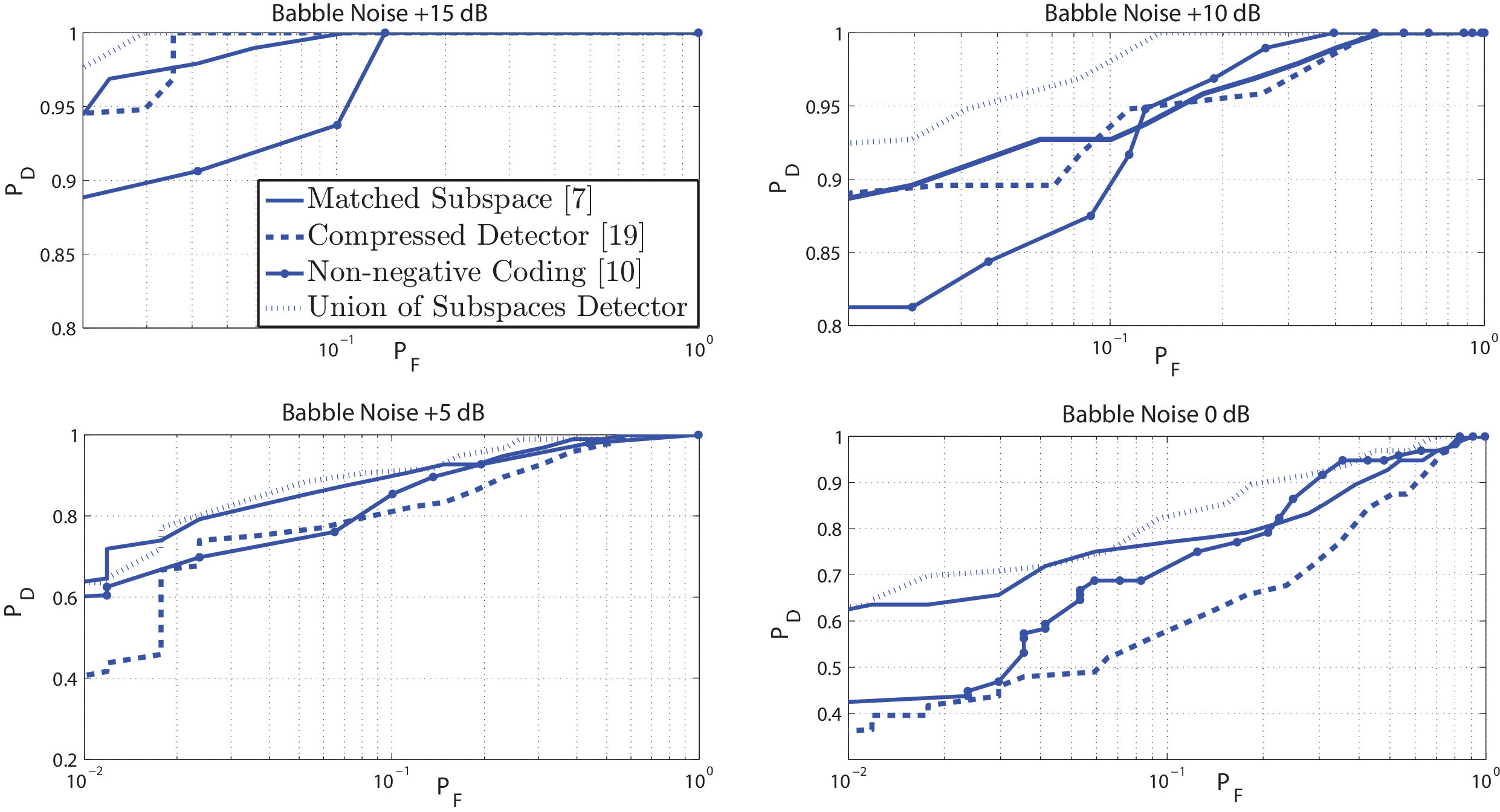}
  \caption{Comparison the performance of our proposed method with sparse non-negative coding based VAD \cite{Teng24}, matched subspace detector \cite{Scharf12} where the bases are found by PCA and detection using the compressed measurements \cite{Davenport06}.}
  \label{simul2}
\end{figure}

Sparsity in voice activity detection has been exploited already. E.g, a feature extraction is performed to suggest a decision rule for detection in \cite{Teng24}.  We compared the result of our method with the sparsity-based VAD method proposed in \cite{Teng24}. As can be seen in Fig. \ref{simul2}, our method shows better performance in low SNR conditions.
\section{Conclusion}
This paper presented a new sparsity-based detector. The performance of the method was evaluated in a realistic application: voice activity detection in speech signal processing. Our detector proposed a new trade-off for designing detectors by assuming the union of low-rank subspaces model. The trade-off is between the sparsity and the error of union of low-rank subspaces model denoted by ESR. In our detector the number of filter banks is proportional to the size of the dictionary. Appropriate dictionary is able to regularize the sparsity and the introduced parameter ESR. Simulation results showed that the proposed method is effective and has a high anti-noise ability due to optimum projection of signals to reliable learned low-rank subspaces.

\bibliographystyle{ieeetr}
\bibliography{ULSD_ref}

\end{document}